\documentclass[twocolumn,twoside]{revtex4}
\usepackage{graphicx}
\usepackage{fancyhdr}
\usepackage{amsmath}
\begin{document}

\title{Hyperon physics at BESIII}


\author{Liang~Yan\\
on behalf of the BESIII Collaboration}
\affiliation{Fudan University, Shanghai, People's Republic of China, 200433}
\affiliation{Key Laboratory of Nuclear Physics and Ion-beam Application (MOE) and Institute of Modern Physics, Fudan University, Shanghai, People's Republic of China, 200433}

\begin{abstract}
Using the largest $J/\psi$ and $\psi(3686)$ data samples collected at the BESIII experiment, the hyperon anti-hyperon pairs are generated in their quantum entangle system. The polarization of hyperon has been observed, which makes it possible to measure the decay parameters of hyperon and anti-hyperon precisely. 
The CP conservation observables have been estimated which is important to test the Standard Model. In the meanwhile, the branching fractions and angular distributions related to hyperon pair production have been measured. 
\end{abstract}

\maketitle

\thispagestyle{fancy}


\section{Introduction}
To explain the fact that matter is more than antimatter in the evolution of the universe, Sakharov gave a necessary condition for this mechanism: there must be an interaction that breaks the conservation of baryon numbers. Furthermore, CP violation should be existence~\cite{Sakharov:1967dj}. The first CP violation observation is in K meson decay by Cronin and Fitch in 1964~\cite{Christenson:1964fg}. In 2001, the BaBar Collaboration and Belle Collaboration announced that CP violation (indirect CP violation) was found in B meson decay~\cite{Belle:2001qdd, BaBar:2001ags}. Successively, direct CP violation was observed during the rare decay of B meson~\cite{Belle:2004nch, BaBar:2004gyj}. In 2019, the LHCb Collaboration announced that CP violation was first found in the charm decays~\cite{LHCb:2019hro}. 
In the Standard Model, the CKM matrix is a unitary matrix which contains information on the strength of the flavor-changing weak interaction. There are  three Euler angles and one phase angle. This phase introduces the important possibility of CP violation amplitudes~\cite{Kobayashi:1973fv}. 
As the most successful theory in the explanation of CP violation, the known experimental results are consistent with the KM mechanism predictions.
However, this is still not enough to explain the material dominance in the universe~\cite{Peskin:2002mm}. 
To search for new sources of CP violation, hyperon physics has attracted more attention recently.  
Hyperon is the baryon containing one or more strange quarks. 
The spin 1/2 hyperons could be used to study the symmetry through their two-body weak decay.  In the process of hyperon decay to nucleon (proton or neutron) and $\pi$ meson, its decay parameters can be described by a parity conserving (P-wave) and a parity violation (S-wave) amplitudes~\cite{Lee:1957he}:
\begin{equation}
\alpha = \frac{2\mathrm{Re}(S^*P)}{|S|^2 + |P|^{2}}, \beta = \frac{2\mathrm{Im}(S^*P)}{|S|^2 + |P|^{2}}, \gamma = \sqrt{1-\alpha^2}\rm cos\phi
\label{eqn:decay_par}
\end{equation}
where the parameter $\alpha$ can be described as the angular distribution asymmetry of the final state particle (nucleon):
\begin{equation}
dN/d\Omega = \frac{1}{4\pi}( 1 + \alpha\mathbf{P} \cdot \mathbf{\hat{p}})
\label{eqn:eq2}
\end{equation}
where $\mathbf{P}$ is the polarization of hyperon, and $\mathbf{\hat{p}}$ is the unit momentum of the nucleon in its parent's rest frame.
Then by comparing the decay parameters of hyperon and anti-hyperon, CP conservation could be tested.
\begin{equation}
A = \frac{\alpha + \bar{\alpha}}{\alpha - \bar{\alpha}}
\end{equation}
The Standard Model predicts that the hyperon CP violation happens at the level of $10^{-5}$ to $10^{-6}$~\cite{Tandean:2002vy}.
According to Eq.\ref{eqn:eq2},  $\alpha$ (decay parameter) and P (polarization) can not be obtained separately. In the BESIII experiment, the hyperon and anti-hyperon pairs are produced by the annihilation of electron and positron at resonance states ($J/\psi$ and $\psi(3686)$).  This quantum entanglement system provides a controllable and accurate CP conservation test. To describe the hyperon anti-hyperon pairs production, the electromagnetic shape factors are introduced~\cite{Faldt:2017kgy}.
These two form factors can be written as two real parameters $\alpha_{\psi}$ and $\Delta\Phi$, which represent the asymmetry of angular distribution and the phase angle between the two form factors. The observable $\Delta\Phi$ is related to the spin polarization of hyperon and anti-hyperon. If $\Delta\Phi$ is not equal to 0, the polarization of the hyperon will be perpendicular to its generation plane. 
The differential cross-section is given as $d\sigma \propto  {\cal{W}}({\boldsymbol{\xi}}) d{\boldsymbol{\xi}}$, where 
\begin{equation*}
\begin{split}
	{\cal{W}}({\boldsymbol{\xi}})=	 &{\cal{T}}_0({\boldsymbol{\xi}})+{{\alpha_{\psi}}}{\cal{T}}_5({\boldsymbol{\xi}})\\
	+&{{\alpha}{\bar{\alpha}}}\left({\cal{T}}_1({\boldsymbol{\xi}})
+\sqrt{1-\alpha_{\psi}^2}\cos({{\Delta\Phi}}){\cal{T}}_2({\boldsymbol{\xi}})
+{{\alpha_{\psi}}}{\cal{T}}_6({\boldsymbol{\xi}})\right)\\
+&\sqrt{1-\alpha_{\psi}^2}\sin({{\Delta\Phi}})
\left({\alpha}{\cal{T}}_3({\boldsymbol{\xi}})
+\bar{{\alpha}}{\cal{T}}_4({\boldsymbol{\xi}})\right).\label{eq:anglW}
\end{split}
\end{equation*}
and ${\cal{T}}_{i}, (i = 0, 1...6)$ are angular functions dependent on  $\boldsymbol{\xi}$, where $\boldsymbol{\xi}$ represents polar angles of hyperons and polar angles and azimuthal angles of nucleons~\cite{Faldt:2017kgy}.

With the data collected at BESIII, the observation of hyperon polarization opens a new window for testing CP conservation in baryon sector, as it allows for simultaneous production and detection of hyperon and anti-hyperon pair two-body weak decay. The CP symmetry tests are performed in the process of $\Lambda$, $\Sigma$ and $\Xi$ pair production. Furthermore, in the $\Xi$ decay, it is possible to perform three independent CP tests and determine the strong phase and weak phase difference.

\section{$J/\psi \to \Lambda \bar{\Lambda}$}
\begin{figure}[!ht]
    \includegraphics[width=0.9\linewidth]{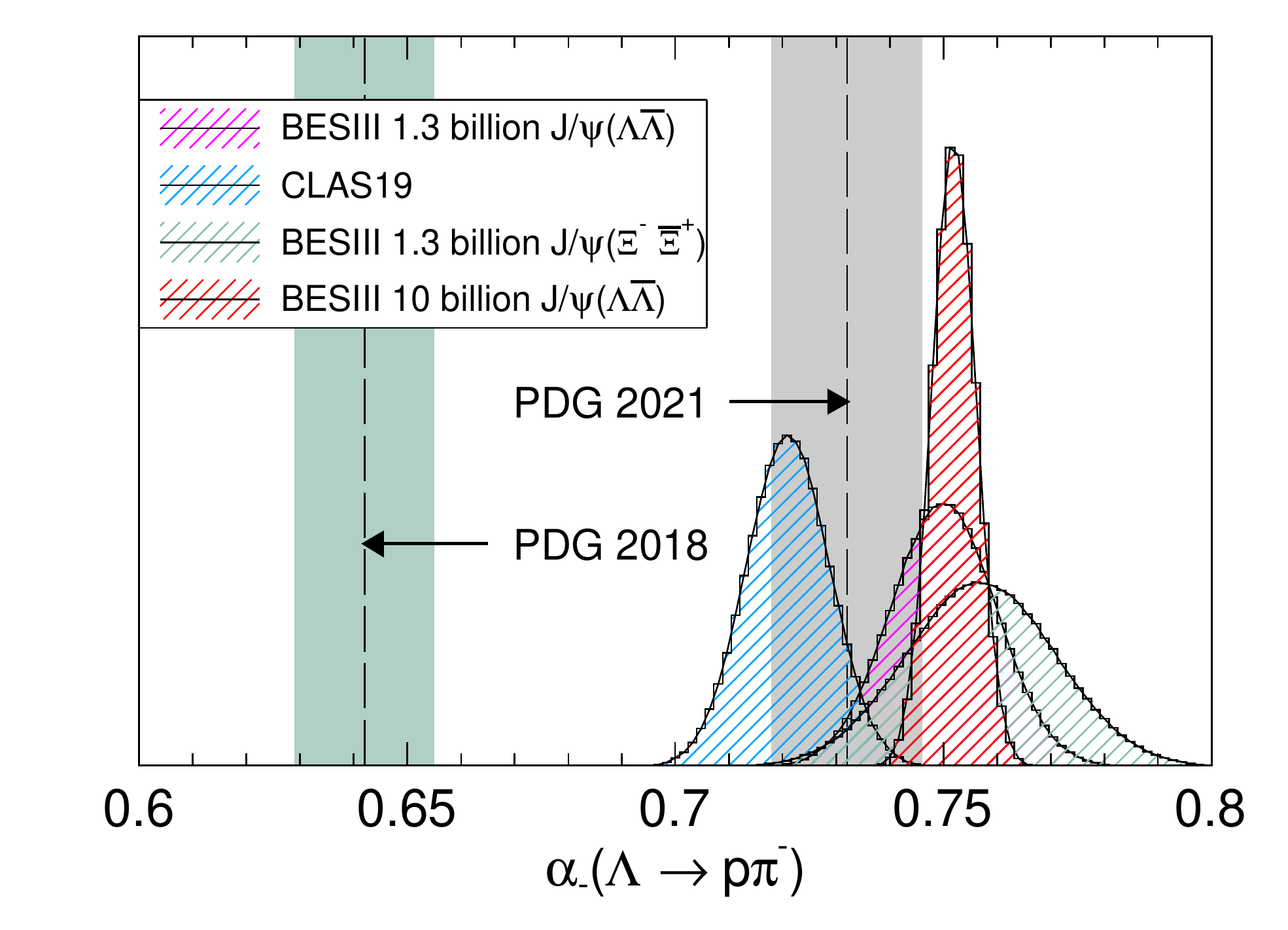} 
    \caption{\label{fig:PDG_value}
    Results of the $\Lambda$ decay parameter from different experiments. 
    The green band represents the PDG 2018 value, 
    and the gray band represents the PDG 2021 value.}
\end{figure}

The process of $J/\psi \rightarrow \Lambda \bar{\Lambda}$ is studied with $1.31 \times 10^9$ $J/\psi$ events collected with BESIII detector~\cite{BESIII:2018cnd}.  
$\Lambda$ is reconstructed with $p \pi^{-}$. $\bar{\Lambda}$ is reconstructed with two modes: $\bar{p} \pi^{+}$ and $\bar{n} \pi^{0}$. In the real data, there are 420,593 and 47,009 signal events are selected for the decay final states $p\pi^{-}\bar{p}\pi^{+}$ and $p\pi^{-}\bar{n}\pi^{0}$ respectively. With the selected signal events, the angular observables ${\boldsymbol{\xi}}$ are fully reconstructed. In the differential cross section formula, the free parameters $\alpha_{\psi}$, $\Delta\Phi$, $\alpha_{-}$, $\alpha_{+}$ and $\bar{\alpha}_{0}$ are extracted by a simultaneous unbinned maximum likelihood fit. The $\Delta\Phi$ is measured to be $(42.4\pm0.6\pm0.5)^\circ$ which means there are a clear polarization existence in the $J/\psi \rightarrow \Lambda \bar{\Lambda}$. Therefore, we could determine the decay asymmetry parameters $\alpha_{-}$, $\alpha_{+}$ and $\bar{\alpha}_{0}$ separately.  The CP violation test $A_{\rm CP} = (\alpha_{-} + \alpha_{+})/(\alpha_{-} - \alpha_{+})$ is measured to be $-0.006 \pm 0.012 \pm 0.007$ which is consistent with SM prediction.  The obtained value of $\alpha_{-}$ is $0.750 \pm 0.009 \pm 0.004$, which is $17\pm3 \%$ higher than the world average at that time. As a fundamental parameter, this measurement has been confirmed by analysis the CLAS data, but with a $2.3\sigma$ descrepancy~\cite{Ireland:2019uja}.  From 2017 to 2019, the BESIII collected 10 billion $J/\psi$ events, which offers the opportunity to improve the $\Lambda$ decay parameters and the CP asymmetry observable with increased precision. The systematic uncertainties are studied in a comprehensive way to match the significant improvement of statistics. As Table~\ref{tab:final_results} shows, the results are consistent with those in the previous analysis with significantly improved precision~\cite{BESIII:2022qax}.
Results of the $\Lambda$ decay parameter from different experiments are shown in Fig.~\ref{fig:PDG_value}. 
The $\alpha_{-}$ value obtained agrees with 
the previous BESIII measurements~\cite{BESIII:2018cnd} and the BESIII result 
extracted from the $J/\psi \rightarrow \Xi^{-}\bar{\Xi}^{+}$ decay~\cite{BESIII:2021ypr}, 
but deviates from the CLAS result by $3.5 \sigma$. 
In addition, we obtain the value of CP violation for the $\Lambda$ decay   
$A_{CP}=(\alpha_{-}+\alpha_{+})/(\alpha_{-}-\alpha_{+})=-0.0025\pm0.0046\pm0.0011$, 
which is compatible with 0, thereby indicating a no-violation scenario.

\begin{table}[h]%The best place to locate the table environment is directly after its first reference in text
\caption{
\label{tab:final_results}
	The angular distribution parameters, $\alpha_{J/\psi}$, $\Delta\Phi$
	and the asymmetry parameters $\alpha_-$ for $\Lambda \rightarrow p\pi^-$, $\alpha_+$
	for $\bar{\Lambda} \rightarrow \bar{p}\pi^+$ obtained in new and
    previous BESIII measurements for comparison.
    The first uncertainty is statistical, the second one is systematic.
    }
\begin{ruledtabular}
\begin{tabular}{ccc}
    \multicolumn{1}{r}{\textrm{Par.}}&
    \multicolumn{1}{c}{\textrm{New results~\cite{BESIII:2022qax}}}&
    \multicolumn{1}{c}{\textrm{Previous results~\cite{BESIII:2018cnd}}}\\
\colrule
$\alpha_{J/\psi}$ &  0.4748 $\pm$ 0.0022 $\pm$ 0.0024 & 0.461 $\pm$ 0.006 $\pm$ 0.007\\
$\Delta\Phi$      &  0.7521 $\pm$ 0.0042 $\pm$ 0.0080 & 0.740 $\pm$ 0.010 $\pm$ 0.009\\
$\alpha_{-}$      &  0.7519 $\pm$ 0.0036 $\pm$ 0.0019 & 0.750 $\pm$ 0.009 $\pm$ 0.004\\
$\alpha_{+}$      &  $-$0.7559 $\pm$ 0.0036 $\pm$ 0.0029 & $-$0.758 $\pm$ 0.010 $\pm$ 0.007\\
\hline
$A_{CP}$          & $-$0.0025 $\pm$ 0.0046 $\pm$ 0.0011 & 0.006 $\pm$ 0.012 $\pm$ 0.007\\
    $\alpha_{\rm{avg}}$    &  0.7542 $\pm$ 0.0010 $\pm$ 0.0020 &  - \\
\end{tabular}
\end{ruledtabular}
\end{table}

\section{$J/\psi$ and $\psi(3686) \rightarrow \Sigma^{+}\bar{\Sigma}^{-}$}
With $1310.6\times10^{6}$ $J/\psi$ and $448.1\times10^{6}$ $\psi(3686)$ events, the studies are performed for $J/\psi$ and $\psi(3686) \rightarrow \Sigma^{+}\bar{\Sigma}^{-}$~\cite{BESIII:2020fqg}. We selected the final states with proton, anti-proton and 2$\pi^{0}$ to reconstruct the signal events. To further suppress the background events, the four momentum kinematic constraints are applied. The peaking background level is below 0.1\%, and other backgrounds are 5\% for $J/\psi$ decay and 1\% for $\psi(3686)$ decay which could be well estimated with sideband method.
An unbinned maximum likelihood fit is performed in the five angular dimensions ${\boldsymbol{\xi}}$, simultaneously fitting both the $J/\psi \rightarrow \Sigma^{+}\bar{\Sigma}^{-}$ and $\psi' \rightarrow \Sigma^{+}\bar{\Sigma}^{-}$ data in order to determine the parameters ${\boldsymbol{\Omega}}=\{\alpha_{J/\psi}, \alpha_{\psi'}, \Delta\Phi_{J/\psi}, \Delta\Phi_{\psi'}, \alpha_{0}, \bar{\alpha}_{0}\}$. 
We could find that $\alpha_{J/\psi}$ is determined to be $-$0.508 $\pm$ 0.006 $\pm$ 0.004 , which is negative as observations made in the decays $J/\psi \to \Sigma^{0}\bar{\Sigma^{0}}$, $J/\psi \to \Sigma(1385)^{-} \bar{\Sigma}(1835)^{+}$ and $J/\psi \to \Sigma(1385)^{+} \bar{\Sigma}(1835)^{-}$. The relative phases $\Delta\Phi_{J/\psi}$ and $\Delta\Phi_{\psi'}$ are determined to be $-0.270\pm0.012\pm0.009$ and $0.379\pm0.07\pm0.014$ simultaneously, which are the first measurements for both reactions $J/\psi \rightarrow \Sigma^{+} \bar{\Sigma}^{-}$ and $\psi' \rightarrow \Sigma^{+} \bar{\Sigma}^{-}$. 
Considering the $\Delta\Phi_{J/\psi}$ and $\Delta\Phi_{\psi'}$ are non-zero, both of them could be used to extract the decay asymmetry parameters $\alpha_{0}$ and $\bar{\alpha}_{0}$ simultaneously. 
The value of $\alpha_{0}$ is determined to be $-0.998\pm0.037\pm0.009$ which is consistent with the PDG average at significantly improved precision, and $\bar{\alpha}_{0}$ is measured to be $0.990\pm0.037\pm0.011$ for the first time. The value of $A_{\rm CP}$ is found to be consistent with CP-conservation and is in agreement with the Standard Model prediction.

\section{$J/\psi \rightarrow \Xi^{-} \bar{\Xi}^{+}$}
\begin{figure}
    \centering
    \includegraphics[width=0.49\textwidth]{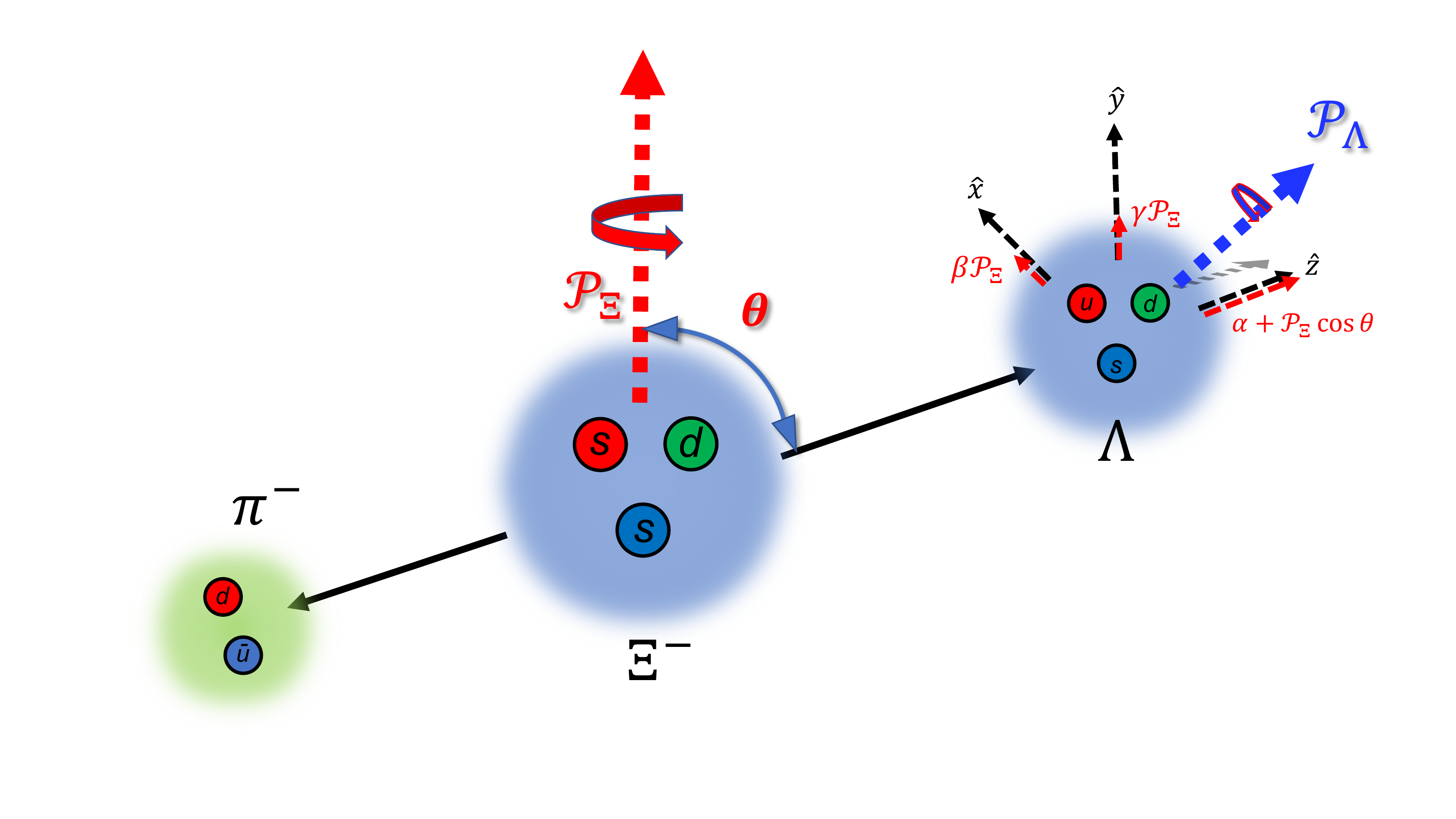}
    \caption{Illustration of the relation between the polarisation vectors of $\Xi$ and $\Lambda$ in terms of the decay parameters $\alpha_\Xi$, $\beta_\Xi$ and $\gamma_\Xi$.}
    \label{fig:decaypars}
\end{figure}

In the previous studies, we only did the measurement of decay parameter $\alpha$. The reason is that the final state polarization of proton could not be determined by the BESIII detectors. 
In the decay process of $\Xi^{-} \rightarrow \Lambda \pi^{-}$, we could also determine the $\alpha_{\Xi}$ as before. The difference is that the polarization of $\Lambda$ could be determined through $\Lambda \rightarrow p \pi^{-}$ which allows us to get all decay parameters: $\alpha_{\Xi}$, $\beta_{\Xi}$ and $\gamma_{\Xi}$. 
As illustrated in Fig.~\ref{fig:decaypars}, the polarizations of $\Xi$ and $\Lambda$ are related to the decay parameters.
Considering these three parameters are not independent as $\alpha_{\Xi}^{2} + \beta_{\Xi}^{2} + \gamma_{\Xi}^{2} = 1$. 
By defining the parameter $\phi_{\Xi}$ according to
\begin{equation}
    \beta_{\Xi} = \sqrt{1-\alpha_{\Xi}^2}\sin\phi_{\Xi}, ~~~\gamma_{\Xi} = \sqrt{1-\alpha_{\Xi}^2}\cos\phi_{\Xi},
    \label{eq:leeyang}
\end{equation}
the decay is completely described by two independent parameters $\alpha_{\Xi}$ and $\phi_{\Xi}$. Then two CP violation observables are defined to be
 \begin{equation}
     A_{\rm CP} = \frac{\alpha_{\Xi}+\overline\alpha_{\Xi}}{\alpha_{\Xi}-\overline\alpha_{\Xi}},  ~~~\Delta \phi_{\rm CP} = \frac{\phi_{\Xi}+\overline\phi_{\Xi}}{2}.
     \label{eq:CPasy}
 \end{equation}
Since the decay amplitude for $\Xi^- \to \Lambda \pi^-$ consists of both a P-wave and an S-wave part, the leading-order contribution to the CP asymmetry $A_{\rm CP}$ can be written as
    \begin{equation}
      A_{\rm CP} \approx - \tan(\delta_P - \delta_S)\tan(\xi_P - \xi_S),
      \label{eq:acpphase}
  \end{equation}
where $\tan(\delta_P-\delta_S) = \beta/\alpha$ denotes the strong phase difference of the final-state interaction between the $\Lambda$ and $\pi^-$ from the $\Xi^{-}$ decay.
An independent CP-symmetry test in $\Xi^-\to\Lambda\pi^-$ is provided by determining the value of $\Delta \phi_{\rm CP}$.
At leading order, this observable is related directly to the weak-phase difference
\begin{equation}
 (\xi_P - \xi_S)_{\rm{LO}} = \frac{\beta + \overline{\beta}}{\alpha - \overline{\alpha}}\approx \frac{\sqrt{1-\alpha^2}}{\alpha}\Delta \phi_{\rm CP}  
 \label{eq:betaprime}
\end{equation} 
and can be measured even if $\delta_P = \delta_S$.

\begin{figure}[h]
    \centering
    \includegraphics[width=0.49\textwidth]{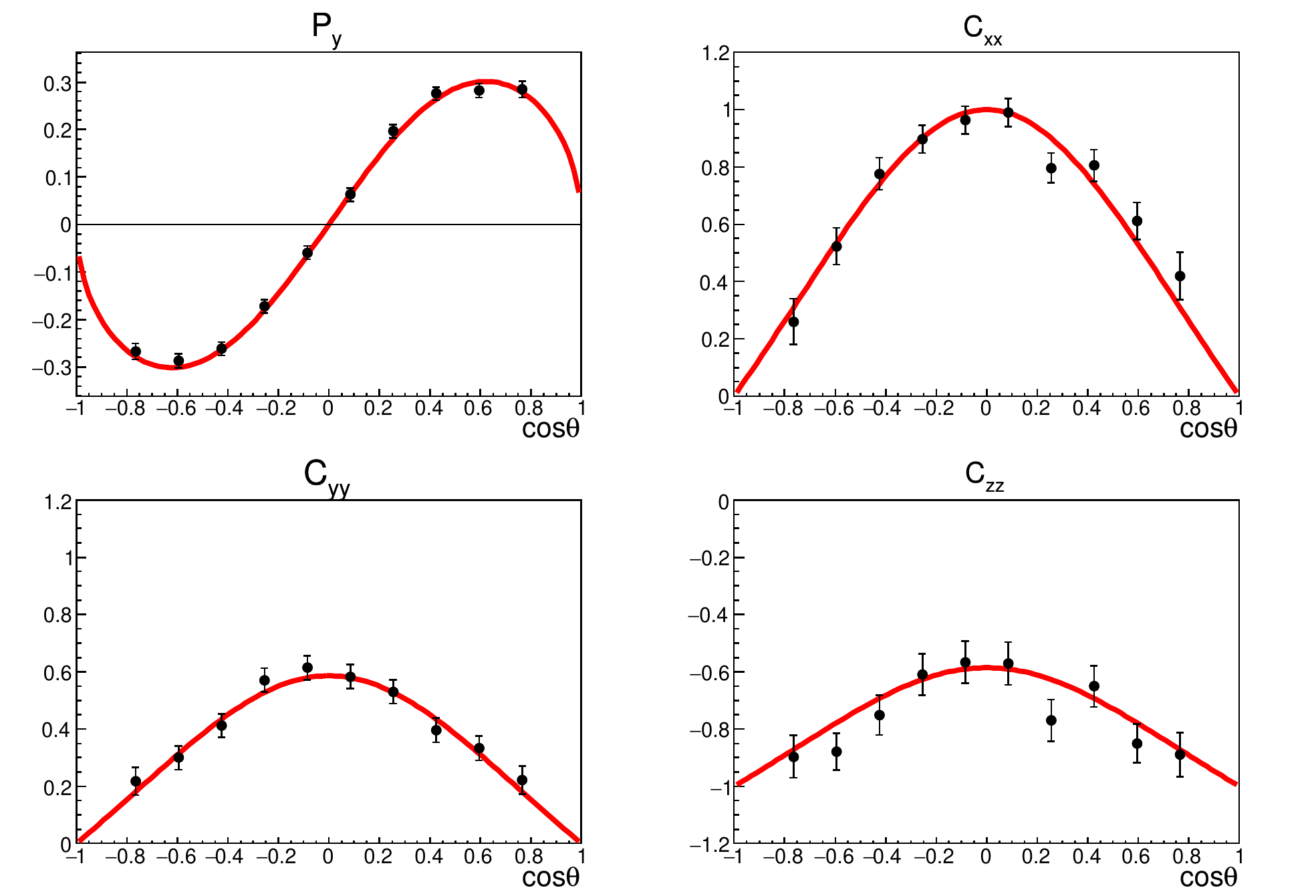}
    \caption{Polarisation and spin correlations and in the $e^+e^-\to\Xi^-\overline\Xi^+$
    reaction. (a) $P_{y}$, (b) $C_{xx}$, (c) $C_{yy}$ and (d) $C_{zz}$ as functions of $\cos\theta$. 
    The data points are determined independently in each bin. The red curves represent the expected angular dependence obtained with the values of $\alpha_{\psi}$ and $\Delta\Phi$ from the global fit. The errors bars indicate the statistical uncertainties.}
    \label{fig:spincorrelations}
\end{figure}

With $1.31\times10^9$ $J/\psi$ events, the multi-dimensional angular fitting is performed to measure the decay parameters in $J/\psi \rightarrow \Xi^{-} \bar{\Xi}^{+}$~\cite{BESIII:2021ypr}. Nine kinematic observables are used to describe this decay process of $J/\psi \rightarrow \Xi^{-} \bar{\Xi}^{+}$, $\Xi^{-} \rightarrow \Lambda \pi^{-}$ and $\bar{\Xi}^{+} \rightarrow \bar{\Lambda} \pi^{+}$. To show the fit quality of the fitting, the polarization and spin correlation plots are shown in Fig.~\ref{fig:spincorrelations}.
Based on the fitting results,
 $A_{\rm CP}$ and  $\Delta \phi_\Xi$ are measured to be $(6.0\pm13.4\pm5.6)\times10^{-3}$ and  $(-4.8\pm13.7\pm2.9)\times10^{-3}~{\rm rad}$, which are consistent with CP symmetry. In addition, the strong and weak $\Xi\to\Lambda\pi$ decay amplitudes could be separated. The direct weak phase difference is measured to be $(\xi_{P} - \xi_{S}) =  (1.2\pm3.4\pm0.8)\times10^{-2}$~rad and is one of the most precise tests of CP symmetry for strange baryons. The strong phase difference is measured to be $(\delta_P - \delta_S) = (-4.0\pm3.3\pm1.7)\times10^{-2}$~rad.

\section{$\psi(3686) \rightarrow \Omega^{-}\bar{\Omega}^{+}$}

The spin of $\Omega^{-}$ is predicted to be 3/2 according to quark model, but it is not unambiguous determined in experiments.
Based on $448.1 \times 10^{6}$ $\psi(3686)$ events collected by BESIII, the process of $\psi(3686) \rightarrow \Omega^{-}\bar{\Omega}^{+}$, with the sequential decays of $\Omega^{-} \rightarrow K^{-} \Lambda, \Lambda \rightarrow p \pi^{-}$ and $\bar{\Omega}^{+} \rightarrow K^{+} \bar{\Lambda}, \bar{\Lambda} \rightarrow \bar{p} \pi^{+}$ has been studied~\cite{BESIII:2020lkm}. To increase the statistics, a single-tag method is implemented in which only the $\Omega^-$ or the $\bar{\Omega}^+$ is reconstructed
via $\Omega^-\to K^-\Lambda\to K^-p\pi^-$ or $\bar{\Omega}^+\to
K^+\bar{\Lambda}\to K^+\bar{p}\pi^+$, and the $\bar{\Omega}^+$ or
$\Omega^-$ on the recoil side is inferred from the missing mass of
the center mass system. With different spin hypotheses (3/2 or 1/2), an unbinned maximum likelihood fit is performed to measure the free parameters with multi-dimensional angular distributions.
By comparing the likelihood values with two assumptions to the around 4000 selected signal events, we determine the significance of the J = 3/2 hypothesis over the J = 1/2 to be larger than 14$\sigma$. 
For the process of $\psi(3868) \rightarrow \Omega^{-}\bar{\Omega}^{+}$, we not only observe vector polarization, but also quadrupole and octupole polarization contributions as the predictions~\cite{Korner:1976hv, Perotti:2018wxm}.

\section{Summary}
With the world's largest $J/\psi$ and $\psi(3686)$ data samples produced at that time, $J/\psi$ and $\psi(3686)$
decays into light hyperon pairs have been studied.  
By exploiting the quantum entangled system, the asymmetry decay parameters are well measured for the hyperon and anti-hyperon.
The CP conservation has been tested with high precision, while the results are consistent with the Standard Model predictions.
The next generation of charm factories will greatly improve the statistics and benefit from these studies, which has potential in CP violation signal observation in baryon sector.

%\begin{acknowledgments}
%\end{acknowledgments}

\bigskip % extra skip inserted
%\begin{thebibliography}{9}   % Use for  1-9  references

\end{document}